\documentclass[preprint,12pt]{elsarticle}

\journal{Physica A: Statistical Mechanics and its Applications}
\usepackage[margin=1in]{geometry}

\usepackage{algorithm}
\usepackage{algpseudocode}
\usepackage{amsbsy, amsfonts, amsmath, amssymb}

\usepackage{soul} 
\usepackage{graphicx}
\usepackage[colorlinks=true, citecolor=blue]{hyperref}
\usepackage{mathrsfs}
\usepackage{bm}
\usepackage{multirow}
\usepackage{booktabs, threeparttable}
\usepackage{setspace}
\usepackage{tikz}
\usepackage{caption} 
\usepackage{diagbox}
\usepackage{subfigure}
\usepackage{longtable}

\usepackage{natbib}

\usepackage{physics} 

\usepackage[inline]{enumitem}

\usepackage[pagewise]{lineno}

\allowdisplaybreaks

\usepackage{xcolor}

\usepackage{graphicx}

\usepackage{tikz}
\usetikzlibrary {matrix}

\usepackage[colorlinks=true, citecolor=blue]{hyperref}
\usepackage{mathrsfs}

\newcommand{\pkg}[1]{{\normalfont\fontseries{b}\selectfont #1}}
\let\proglang=\textsf \let\code=\texttt

\newcommand{\bfW}{\bm{W}}
\newcommand{\bfA}{\bm{A}}
\newcommand{\outdeg}{d^{\, \rm (out)}}
\newcommand{\indeg}{d^{\, \rm (in)}}
\newcommand{\outs}{s^{\, \rm (out)}}
\newcommand{\ins}{s^{\, \rm (in)}}
\newcommand{\PR}{{\rm PR}}
\newcommand{\bfbeta}{\bm{\beta}}
\newcommand{\bfM}{\bm{M}}
\newcommand{\bfP}{\bm{P}}
\newcommand{\bfB}{\bm{B}}

\graphicspath{{../figures/}}

\begin{document}
	
	\begin{frontmatter}
		
		
		
		\title{PageRank centrality and algorithms for weighted, 
		directed networks with applications to World Input-Output 
		Tables}
		
		
		\author[label1]{Panpan Zhang}
		
		\affiliation[label1]{organization={Department of 
		Biostatistics, Epidemiology and	Informatics, University 
		of Pennsylvania},
			city={Philadelphia},
			postcode={19104}, 
			state={PA},
			country={USA}}
		
		\author[label2]{Tiandong Wang}
		
		\affiliation[label2]{organization={Department of Statistics, 
		Texas A\&M University},
			city={College
				Station},
			postcode={77843}, 
			state={TX},
			country={USA}}
		
		\author[label3]{Jun Yan}
		
		\affiliation[label3]{organization={Department of Statistics, 
		University of Connecticut},
			city={Storrs},
			postcode={06269}, 
			state={CT},
			country={USA}}

		
		
		
		\begin{abstract}
			PageRank (PR) is a fundamental tool for assessing the 
			relative 
			importance of the nodes in a network. In this paper, we
			propose a measure, weighted PageRank (WPR), extended 
			from the
			classical PR for weighted, directed networks with 
			possible
			non-uniform node-specific information that is dependent 
			or
			independent of network structure. A tuning parameter 
			leveraging
			node degree and strength is introduced.	An efficient 
			algorithm 
			based on \proglang{R} program has been developed for 
			computing 
			WPR in large-scale networks. We have tested the proposed 
			WPR on 
			widely used simulated network models, and found it 
			outperformed 
			other competing measures in the literature. By applying 
			the 
			proposed WPR to the real network data generated from 
			World 
			Input-Output Tables, we have seen the results that are 
			consistent with the global economic trends, which 
			renders it a 
			preferred measure in the analysis.
			
		\end{abstract}
		
		
		
		
		\begin{keyword}
			node centrality \sep weighted directed networks \sep 
			weighted PageRank \sep World Input-Output Tables
			
			
			\MSC[2008] 91D03 \sep 05C82
		\end{keyword}
		
	\end{frontmatter}

\section{Introduction}
\label{sec:intro}

Centrality measures are widely accepted tools for assessing the
relative importance of the entities in networks. A variety of
centrality measures have been developed in the literature,
including position/degree centrality \citep{freeman1978centrality},
closeness centrality~\citep{newman2001scientific}, betweenness
centrality~\citep{newman2001scientific}, eigenvector
centrality~\citep{bonacich1972factoring}, Katz
centrality~\citep{katz1953anew}, and
PageRank~\citep{brin1999thepagerank}, among others.
Centrality measures have been applied to different types of real
networks; for instance, ranking the cities with at least one
operating airport in the air transport network of China
\citep{wang2011exploring} and evaluating the impact of research
papers in a citation network \citep{hirsch2005index}.
See~\citet{das2018study} for a concise review
and~\citet[Chapter~7]{newman2018networks} for a text-style
elaboration.

The classical PageRank~\citep[PR,][]{brin1998theanatomy} was
designed to precisely rank web pages in Google search via
hyper-textual information (primarily link structure). Today, PR
and its extensions are popular tools for the analyses of all kinds of
networks, such as co-authorship networks~\citep{yan2011discovering},
citation networks~\citep{ding2011applying}, and biological
networks~\citep{kalecky2018primalign}. One limitation of the classical
PR is that it does not account for edge weight in definition. Although
ignoring edge weight may sometimes help with a quick exploration of 
the
fundamental structure of a network, the discarded edge weight can lead
to incorrect inference \citep{newman2004analysis}.
Only a limited number of works considered weight for PR.
\citet{xing2004weighted} asserted that the popularity of a web page
should be based on the numbers of its in-links and
out-links. \citet{ding2011applying} suggested replacing the random
restart of the new process with a probability distribution based on
the weights assigned to the nodes. No PR centrality measures have
put edge weight and node weight in a unified framework.

As most real networks are weighted and directed, possibly with 
node-specific auxiliary information,
here we consider a weighted PageRank (WPR) measure.
The WPR uses edge direction and weight as well as auxiliary
information at the node level to better characterize the centrality 
of a network. The computation of WPR boils down to finding the
principal eigenvector of a big matrix, which can be efficiently done
for large networks. We assess the performance of the proposed WPR in 
comparison with
a few extended PR measures in the literature though numerical studies
with synthetic networks. In the applications to the
World Input-Output networks (WIONs) constructed from World 
Input-Output
Tables \citep[WIOTs,][]{timmer2015anillustrated}, the proposed WPR 
measure
gives more intuitive results than the existing PR measures.
The implementation of the proposed and competing
measures is publicly available in an open-source \proglang{R}
package \proglang{wdnet} \citep{Rpkg:wdnet}.

The rest of the manuscript is organized as follows. In
Section~\ref{sec:WPR}, we propose a WPR measure and demonstrate the
computation strategy. We carry out some synthetic data analyses in
Section~\ref{sec:sim}, where two classes of widely used network
models, namely scale-free networks and stochastic block models, are
adopted. We apply the proposed WRP measure to the WIONs in
Section~\ref{sec:wion}, followed by some concluding remarks and
discussions in Section~\ref{sec:dis}.

\section{Weighted PageRank}
\label{sec:WPR}

We begin with some basic network notations. Let $G(V, E)$
denote a weighted and directed network, where $V$ and $E$ are
respectively its node and edge sets. The structure of $G$ is
characterized by its weighted adjacency matrix $\bfW := (w_{ij})$,
where $w_{ij}$ is the weight of the directed edge from $i \in V$ to
$j \in V$. If no edge exists from $i$ to $j$, then $w_{ij} = 0$.
When edge weight is ignored, $\bfW$ is reduced to the
standard adjacency matrix $\bfA := (a_{ij})$, where $a_{ij}$ takes
value~$0$ or~$1$. For any $i \in V$, let $\outdeg_i := \sum_{j \in
V} a_{ij}$ and $\indeg_i := \sum_{j \in V} a_{ji}$ respectively
denote the out-degree and in-degree of $i$, referring to the numbers
of edges emanating out from and pointing into $i$. Analogously, we
have $\outs_i := \sum_{j \in V} w_{ij}$ and
$\ins_i := \sum_{j \in  V} w_{ji}$,
respectively called the out-strength and in-strength of $i$
when weight is accounted.

\subsection{Formulation}
\citet{brin1998theanatomy} defined PR recursively as
\begin{equation}
	\label{eq:PR}
	\PR(i) = \gamma \sum_{j \in V} \frac{a_{ji}}{\outdeg_j} \PR(j) +
	\frac{1 - \gamma}{n},
\end{equation}
where $\PR(i)$ is the PR of node $i$,
$n = |V|$ counts the number of nodes in $G(V, E$), and
$\gamma \in [0, 1)$ is a damping factor ensuring the algorithm
never gets stuck in a ``sinking node''. This definition is
is based on a random surfer model.
Suppose that an Internet surfer keeps clicking
on links bringing her to different web pages. With probability
$(1 - \gamma)$ she restarts the
process by randomly selecting a web page as the new initial state,
where $\gamma$ is the probability that she continues
in the current process. The inclusion of damping factor in the model
ensures that the process will not be forced to terminate when the
surfer arrives at a web page with no outbound link, called sinking
node. Equation~\eqref{eq:PR} suggests that a node would get a high
PR score if: (1) it receives a large number of incoming edges; (2)
senders of those incoming edges have small out-degrees; or (3) the
PR scores of the senders are high.

In practice, lots of real networks are directed, weighted, and 
affiliated with
important node-specific information.
For instance, in Section~\ref{sec:wion}, we consider the WIONs whose 
nodes correspond to different region-sectors, and
edge weights are determined by transaction volumes.
In addition, the total value added for each region-sector is considered as 
node-specific information.
Here we extend the
classical PR to a weighted version by simultaneously considering
edge weights and auxiliary information contained in nodes.
Let $\phi(i)$ denote the weighted PR of $i$, and $\beta_i$ be
some node-specific quantifiable information attached to $i$. We
assume that $\beta_i$ is independent of $w_{ij}$ for all
$i, j \in V$.

Analogous to Equation~\eqref{eq:PR}, we define the weighted PR 
 recursively by
\begin{equation}
	\label{eq:WPR}
	\phi(i) = \gamma \sum_{j \in V} \left(\theta
	\frac{w_{ji}}{\outs_j} + (1 -
	\theta)\frac{a_{ji}}{\outdeg_j}\right)\phi(j) + \frac{(1 -
	\gamma) \beta_i}{\sum_{i \in V} \beta_i},
\end{equation}
where $\theta \in [0, 1]$ is a tuning parameter adjusting the
relative importance of weights in the definition. The value of
$\theta$ can be chosen according to practical needs and actual
interpretations. For instance, the value of a business project may
be heavily reflected in the investment amount it has received rather
than the number of investors, and
the popularity of a product mainly depends on the sales volume. In
many situations, a balance between the two factors is needed. For
example, the strength of a researcher is related to the number of
publications as well as the prestige of the journals (measured by a
unified metric such as impact factor) of the publications
simultaneously. The tuning parameter $\theta$ controls the balance between
weight and degree. For the special case of $\theta = 0$, the 
proposed WPR is equivalent to the weighted PR introduced 
by~\citet{ding2011applying}. The vector
$\bfbeta := (\beta_1, \beta_2, \ldots, \beta_n)^{\top}$
usually takes the non-uniform relative importance of the nodes into 
account. When no such
information is available, we let $\beta_i = 1$, $i=1,\ldots,n$, so 
that the second term on the right-hand-side (RHS) of
Equation~\eqref{eq:WPR} coincides with what has been defined in
Equation~\eqref{eq:PR}. 

\subsection{Computation}
We propose an efficient algorithm for the computation of the 
proposed WPR in large networks.
The standard method to compute classical PR is the power
iteration. It is well known that the power iteration converges
slowly, especially for massive and dense networks. This leads to the
development of accelerated algorithms, many of which have been
surveyed in~\citet{berkhin2005asurvey}. 
When $\gamma\neq 1$, the underlying process of the random
surfer model for the classical PR is a irreducible Markov
chain~\citep{berkhin2005asurvey}, where
every state in the chain can
be accessed with positive probability from other states.
Here we regard nodes in a network as
states in a Markov chain.

A similar argument can be applied
to the proposed WPR. Let $\bfM := (m_{ij})$ be the transition matrix
of the associated Markov chain for WPR, where
\[
m_{ij} =
\begin{cases}
	\theta w_{ji} / \outs_j + (1 - \theta) a_{ji} /
	\outdeg_j, &\qquad \text{if } \outdeg_j \neq 0;
	\\ \beta_i / \sum_{i \in V} \beta_i, &\qquad \text{if }
	\outdeg_j = 0.
\end{cases}
\]
Notice that
\[\sum_{i \in V} \left(\theta \frac{w_{ji}}{\outs_j} + (1 -
\theta)\frac{a_{ji}}{\outdeg_j}\right) = \theta + (1 - \theta) = 1.\]
That is, matrix $\bfM$ is non-negative and column stochastic. Let
$\bfP := \bigl(\phi(1), \phi(2), \ldots, \phi(n)\bigr)^{\top}$
be a column vector collecting the WPR for each node.
Equation~\eqref{eq:WPR} is equivalent to
\begin{equation}
	\label{eq:WPR2}
	\bfP = \gamma \bfM \bfP + (1 - \gamma)\bfbeta^*,
\end{equation}
where $\bfbeta^* = \bfbeta / \| \bfbeta \|_1$ is the normalization
of $\bfbeta$.

Since $\bfM$ is column stochastic, we can
rewrite Equation~\eqref{eq:WPR2} as
\begin{equation}
	\label{eq:WPR3}
	\bfP = \left(\gamma \bfM + (1 - \gamma) \bfB\right)\bfP =: \bfM^*
	\bfP,
\end{equation}
where $\bfB$ is an $(n \times n)$ matrix such that the $i$-th column
is given by $\beta^*_i \cdot {\bf 1}$ for $i = 1, 2, \ldots, n$.
It is obvious that $\bfB$ is also column stochastic, rendering that
$\bfM^*$ is strictly positive and column stochastic provided that
$\bfbeta \neq {\bf 0}$. By the Perron-Frobenius
theorem~\citep{perron1907zur}, the largest eigenvalue of $\bfM^*$ is
equal to~$1$, and the solution to Equation~\eqref{eq:WPR3} is the
corresponding eigenvector. In the context of stochastic process, we
regard the normalized solution of $\bfP$ as a stationary
distribution of the Markov chain associated with the probability
transition matrix $\bfM^*$. Based on the above representation, the
computation of $\bfP$ in a massive network is converted to finding
the principal eigenvector of a large-scale matrix. One of the most
efficient approaches is the \proglang{ARPACK} software
\citep{lehoucq1998arpack}, with a recent interface for \proglang{R}
through package \pkg{rARPACK} \citep{qiu2016rARPACK}.

\section{Synthetic Data Examples}
\label{sec:sim}

We assess the performance of the proposed WPR measure with
scale-free networks and stochastic block networks. Comparison is done
with respect to classical PR~\citep{brin1999thepagerank} and two existing
weighted PR measures respectively introduced
by~\citet{xing2004weighted} and~\citet{ding2011applying}. The
definition of the former is relegated Appendix~\ref{sec:xing}, 
where the latter is a special case of the proposed WPR as mentioned.
Under each network model, we introduced a parameter
$\rho \in [0, 1]$ to control the strength of the dependency between
the node-specific prior information and the node strength.
Specifically, let $S_i$ be the strength of node~$i$, $i \in V$. Then,
the prior information $\beta_i$ of this node was generated such that
the correlation between $S_i$ and $\beta_i$ is $\rho \in [0, 1]$. This
can be done by setting $\beta_i = \alpha S_i + (1 - \alpha) X_i$,
where $X_i$ is a postive random variable independent of $S_i$ and
\begin{equation}
  \label{eq:linearmix}
  \alpha = \left(1 + \sqrt{\frac{(1 - \rho^2){\rm Var}(S_i)}{\rho^2
        {\rm Var}(X_i)}}\right)^{-1}.
\end{equation}
As $\rho \to 0$, we have $\beta_i \to X_i$, which coincides with the 
case
of~\citet{ding2011applying}. When $\rho = 1$, $\beta_i = S_i$.
For any $\rho \in (0, 1)$, $\beta_i$'s generated with
Equation~\eqref{eq:linearmix} were used as prior information for
WPR computations.

\subsection{Scale-free Network}
\label{sec:sf}

\begin{figure}[tbp]
	\begin{center}
		\includegraphics[width = 0.68\textwidth]{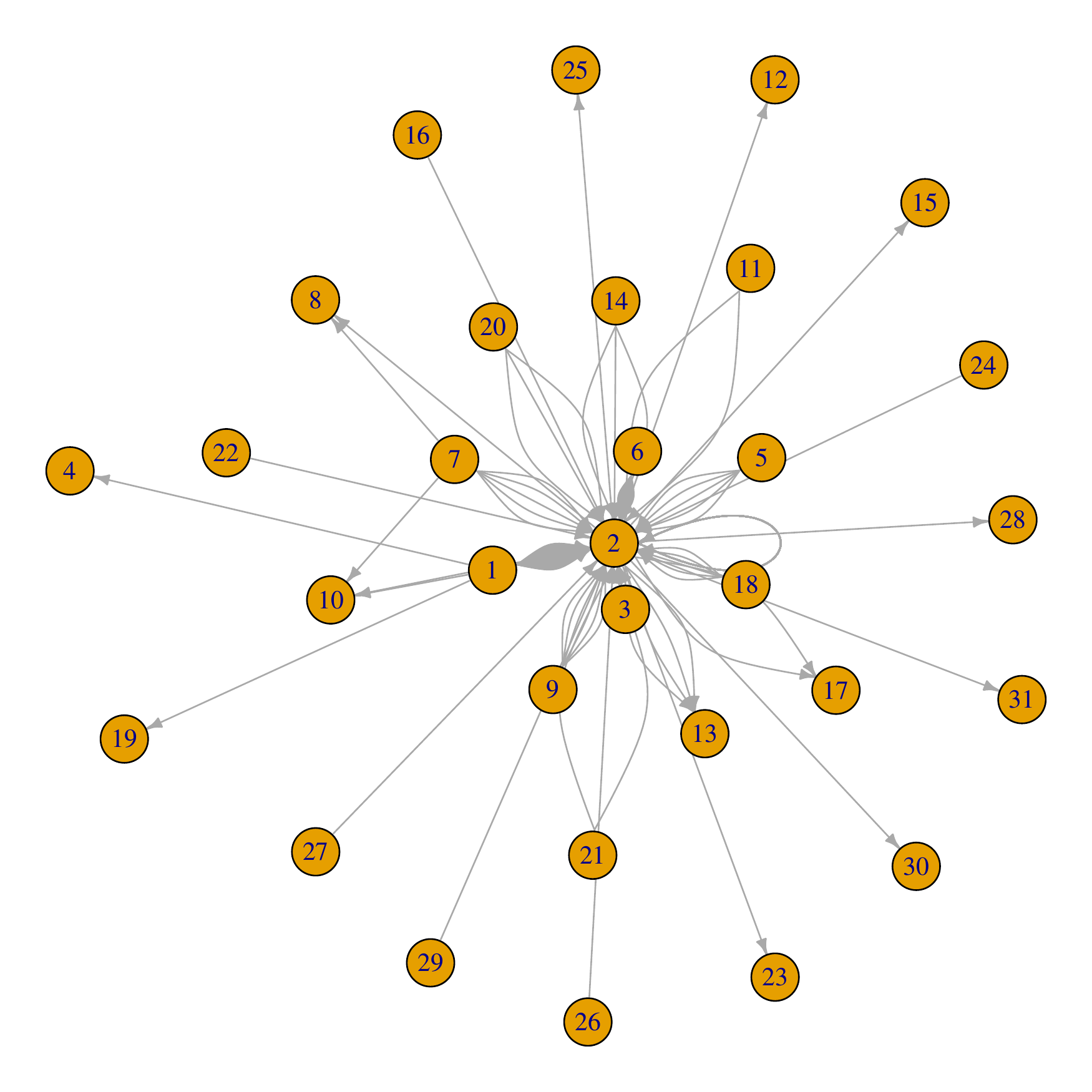}
	\end{center}
	\caption{A simulated (weighted and directed) PA network
	mimicking a Facebook wall post data.}
	\label{fig:PA_ex}
\end{figure}

In the literature, the preferential
attachment (PA) rule~\citep{barabasi1999emergence} is one way
to generate scale-free networks. We used
the algorithm of~\citet{yuan2021assortativity} to generate weighted
directed PA networks via \proglang{R} package 
\code{wdnet}~\citep{Rpkg:wdnet}. Specifically, the simulated PA 
network initiates 
with a directed edge from node $1$ to $2$, where the weight is drawn 
from ${\rm Bin}(100, 0.75)$. At each subsequent step, an edge is 
added according to one of the following three scenarios: With 
probability $\alpha = 0.05$, the edge is added from a new node to an 
existing one; with probability $\beta = 0.9$, the edge is added 
between two existing nodes; with probability $\gamma = 0.05$, the 
edge is added from an existing node to a new one. The source and 
target nodes of the added edge are selected proportional to their 
current out- and in-strengths, respectively. Upon the edge being 
added, its weight is independently drawn from ${\rm Bin}(100, 
0.75)$. The leveraging parameters $\delta_{\, \rm in}$ and 
$\delta_{\, \rm out}$ that respectively control the growth rates of 
in-strengths and out-strengths are fixed, both taking value $1$. We 
refer the readers to~\citet{wang2021directed} for the statistical 
properties and detailed interpretations of these parameters. The 
evolution proceeds in this fashion for $300$ steps, where the 
resulting PA network is depicted in Figure~\ref{fig:PA_ex}.

\begin{table}[tbp]
	\centering
	\caption{A comparison of the proposed WPR measure with those
		respectively proposed in \citet{ding2011applying} 
		(equivalent to the case of $\theta = 0$) and
		\citet{xing2004weighted} for the simulated PA network; the
		damping factor for the proposed WPR is fixed $\gamma =
		0.85$.}
	\label{tab:PA_sim}
	\setlength{\tabcolsep}{7pt}
	\begin{tabular}{cc cc cc cc}
		\toprule
		\multicolumn{2}{c}{$\phi$ ($\theta = 0)$} &
		\multicolumn{2}{c}{$\phi$ ($\theta = 0.5$)} &
		\multicolumn{2}{c}{$\phi$ ($\theta = 1$)} &
		\multicolumn{2}{c}{X-G's measure} \\
		\cmidrule(lr){1-2} \cmidrule(lr){3-4}
		\cmidrule(lr){5-6} \cmidrule(lr){7-8}
		Node & WPR (\%) & Node & WPR (\%) & Node & WPR (\%) & Node & 
		WPR (\%) \\
		\midrule
		2 & 26.216 & 2 & 38.032 & 2 & 66.030 & 3 & 1.710 \\
		10 & 4.415 & 10 & 3.404 & 13 & 2.177 & 5 & 1.710 \\
		8 & 4.051 & 8 & 3.225 & 17 & 1.643 & 6 & 1.710 \\
		13 & 3.567 & 13 & 3.185 & 10 & 1.429 & 9 & 1.710 \\
		31 & 3.567 & 17 & 3.031 & 8 & 1.377 & 11 & 1.710 \\
		17 & 3.567 & 28 & 2.931 & 28 & 1.294 & 14 & 1.710 \\
		25 & 3.567 & 23 & 2.928 & 23 & 1.286 & 16 & 1.710 \\
		28 & 3.567 & 30 & 2.928 & 30 & 1.286 & 18 & 1.710 \\
		30 & 3.567 & 25 & 2.927 & 25 & 1.282 & 20 & 1.710 \\
		12 & 3.567 & 15 & 2.925 & 15 & 1.273 & 21 & 1.710 \\
		\bottomrule
	\end{tabular}
\end{table}

In the first experiment, we did not consider any kind of 
node-specific
prior information, that is, $\beta_i = 1$ for all $i \in V$.
Table~\ref{tab:PA_sim} summarizes the top~10 nodes based on Xing and 
Ghorbanis' PR measure and the proposed WPR measures (with $\gamma = 
0.85$ and $\theta = \{0, 0.5, 1\}$)
With Xing and Ghorbanis' PR measure, all of the top 10 nodes
have the same score, so they are simply ordered by their appearance
timing. This does not provide much practical guidance.
In particular, node~2, which emerges at the central
position in Figure~\ref{fig:PA_ex}, does not appear in the top~10
list. Hence, Xing and Ghorbanis' PR measure will not be
considered in the sequel.

Node~2 ranks first in all three lists in Table~\ref{tab:PA_sim}, which
is consistent with the observation from Figure~\ref{fig:PA_ex}. The
nodes of rank~2
and 3 are the same for $\theta = 0$ and $\theta = 0.5$, but not for
$\theta = 1$. Ding's PR measure ($\theta = 0$) could not distinguish the
nodes from rank~4 to~10, as they have exactly the same score.
This is expected; a measure not accounting for edge weight is not
suitable for weighted networks. From the list of $\theta =0.5$,
nodes of lower ranks (in top~10)
have become identifiable, albeit with tiny gaps. When edge weights are
fully accounted, node~2 is extensively dominant in
the network with a much higher WPR score than the rest.
Meanwhile, the normalized WPR score of node~2 in the list of
$\theta = 1.0$ is higher than the counterparts in the lists of
$\theta = 0$ and $\theta = 0.5$ as well. The ranks of nodes~13 and~17
both rise in the list of $\theta = 1.0$, while those of nodes~10 and~8
drop.  Further investigation reveals that node~13 and~17 both have
links of high weight 295 and 165, respectively, from node~2.
Except for edges pointing towards node 2, these two
edges are the most weighted in the network. Though node 10 receives
three links respectively with weight 72 from node~1, weight 80 from
node~2, and weight 72 from node~7, and node~8 receives two links
respectively with weight 71 from node~2 and weight 76 from node 7,
the incoming edges from node~2 are relatively small and the WPR
scores of all of the other nodes are much smaller than that of
node~2. As a result, nodes~10 and 8~are ranked lower than nodes~13
and~17.

\begin{table}[tbp]
	\centering
	\caption{Nodes of top 10 WPR scores with/without (independent)
	prior information in the simulated PA network for $\theta \in
	\{0, 0.5, 1\}$ and $\gamma =
	0.85$.}
	\label{tab:PA_sim_prior}
	\setlength{\tabcolsep}{7pt}
	\begin{tabular}{cccccc}
		\toprule
		\multicolumn{2}{c}{$\theta = 0$} &
		\multicolumn{2}{c}{$\theta = 0.5$} &
		\multicolumn{2}{c}{$\theta = 1$} \\
		\cmidrule(lr){1-2} \cmidrule(lr){3-4}
		\cmidrule(lr){5-6}
		no prior & with prior & no prior & with prior & no prior &
		with prior \\
		\midrule
		2 & 2 & 2 & 2 & 2 & 2 \\
		10 & 19 & 10 & 19 & 13 & 19 \\
		8 & 10 & 8 & 10 & 17 & 7 \\
		13 & 8 & 13 & 7 & 10 & 10 \\
		31 & 7 & 17 & 17 & 8 & 4 \\
		17 & 17 & 28 & 4 & 28 & 17 \\
		25 & 4 & 23 & 8 & 23 & 29 \\
		28 & 23 & 30 & 23 & 30 & 23 \\
		30 & 25 & 25 & 25 & 25 & 25 \\
		12 & 29 & 15 & 29 & 15 & 8 \\
		\bottomrule
	\end{tabular}
\end{table}

\begin{figure}[tbp]
  \begin{center}
    \includegraphics[width = \textwidth]{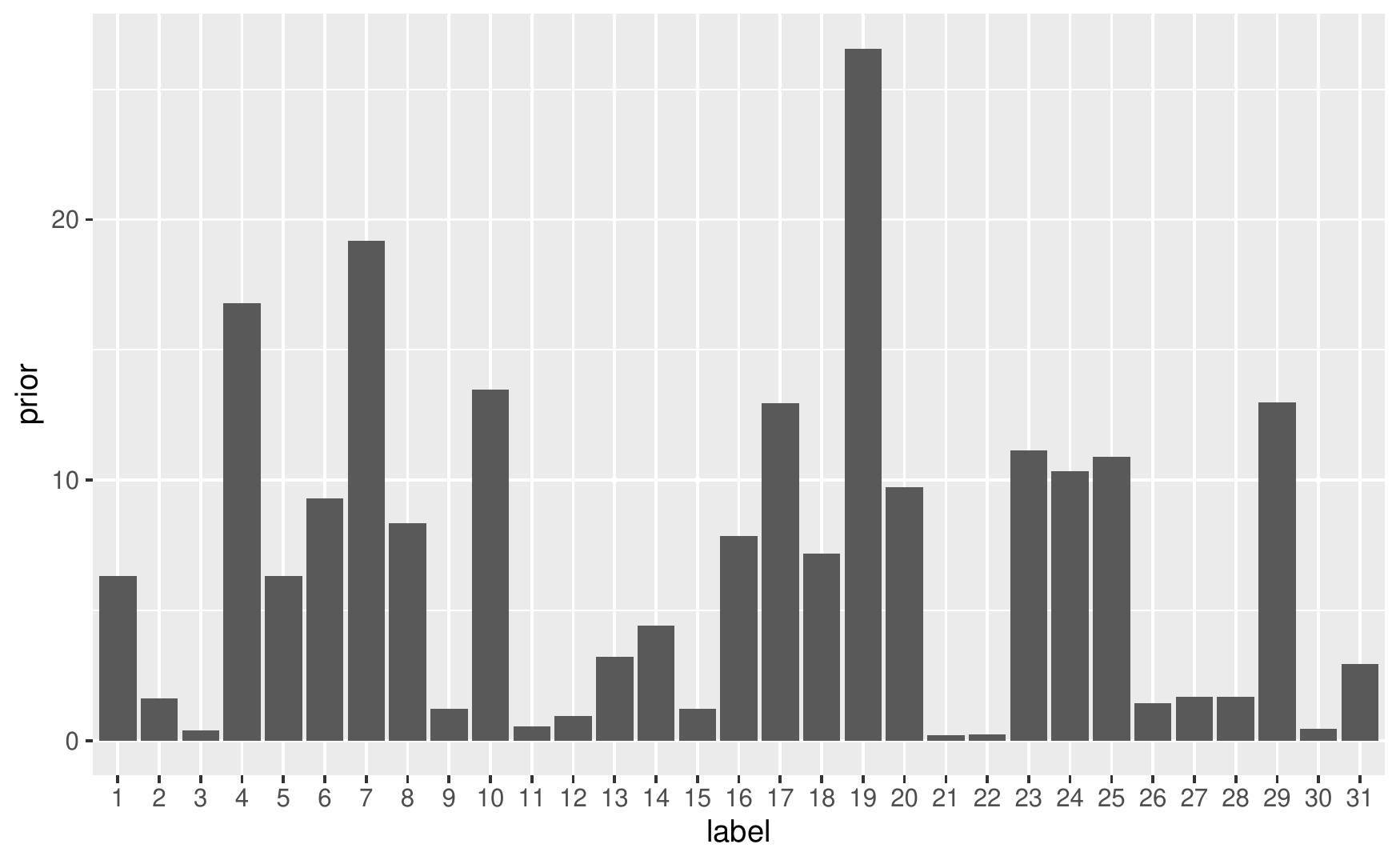}
  \end{center}
  \caption{Generated scores (as prior information) for the nodes
    in the simulated PA network.}
  \label{fig:PA_prior}
\end{figure}

In the second experiment, we incorporated an independent 
node-specific
prior ranking weight ($\rho = 0$) into the WPR computation. The priors
were generated independently from the network with an exponential
distribution with mean~$5$; see Figure~\ref{fig:PA_prior}.
Table~\ref{tab:PA_sim_prior} summarizes 
the top~10 nodes based on
proposed WPRs obtained under $\theta \in \{0, 0.5, 1\}$ and
$\gamma = 0.85$ in comparison to those without considering the prior
information. Drastic changes are observed.
Take $\theta = 1$ as an example. Four nodes in the top~10 list are
new. Two of them, nodes~19 and~7 rank~2 and~3, respectively,
compared to~13 and~31 without the prior information. Indeed, these two
nodes indeed have the top two prior scores as shown in in
Figure~\ref{fig:PA_prior}. Although node~2 remains at
the top, inclusion of the prior information has brought several low
rank nodes to much higher ranks. This suggest that a non-negligible
impact of the prior information on WPR and the ultimate ranking.

Lastly, we carried out a sensitivity analysis for
$\theta \in \{0, 0.5, 1\}$ and $\rho \in \{0.25, 0.5, 0.75\}$, where
$\rho$ is the correlation between the prior information and the
in-strength.  Table~\ref{tab:PA_sim_prior_rho} summarizes the results
of the top~10 nodes. Especially for $\theta = 1$,  the ranks are 
almost identical for different
choices of $\rho$, suggesting that the proposed WPR is robust when
edge weight is fully accounted in the computation. For $\theta =
0.5$, we do not observe significant difference in node labels across
the three lists. The new participant for the list of $\rho = 0.75$, 
node~13, is ranked 12 and 11 respectively in the lists of $\rho = 
0.25$
and $\rho = 0.5$, and node 29 has dropped to rank 11 in the list of
$\rho = 0.75$. There have been some changes in the rank orders as
expected, as the quantities of resulting priors change with the
value of $\rho$. For instance, the prior of node 7 is much higher
than that of node 17 for small $\rho$, but the deviation gets
smaller as $\rho$ gets larger. As node 17 receives a moderately
weighted link from node 2 (the one with the largest WPR
score), it takes the fourth place in the lists of $\rho = 0.5$ and
$\rho = 0.75$. Furthermore, node 8 has surpassed node 7 in the list
of $\rho = 0.75$, too, as it also gets a link from node 2, albeit
its small prior. When weight is not considered, we again only see
difference in the order of ranks in the presented lists. For $\rho =
0.75$, node 10 has replaced node 19 taking the second place. Node 19
is always ranked high since it has the largest prior of all, but
there is only one link pointing to it. The difference between the
priors of node 10 and node 19 is not big for $\rho = 0.75$, but node
10 receives more links from the others in the network, including one
from node 2, rendering it to take a higher rank ultimately.

\begin{table}[tbp]
	\centering
	\caption{Nodes of top 10 WPR scores with a variety of correlated
		prior information, $\rho \in \{0.25, 0.5, 0.75\}$, in the
		simulated PA network for $\theta \in \{0, 0.5, 1\}$ and
		$\gamma = 0.85$.}
	\label{tab:PA_sim_prior_rho}
	\setlength{\tabcolsep}{5.5pt}
	\begin{tabular}{ccccccccc}
		\toprule
		\multicolumn{3}{c}{$\theta = 0$} &
		\multicolumn{3}{c}{$\theta = 0.5$} &
		\multicolumn{3}{c}{$\theta = 1$} \\
		\cmidrule(lr){1-3} \cmidrule(lr){4-6}
		\cmidrule(lr){7-9}
		$\rho = 0.25$ & $\rho = 0.5$ & $\rho = 0.75$ & $\rho = 0.25$
		& $\rho = 0.5$ & $\rho = 0.75$ & $\rho = 0.25$ & $\rho =
		0.5$ & $\rho = 0.75$ \\
		\midrule
		2 & 2 & 2 & 2 & 2 & 2 & 2 & 2 & 2 \\
		19 & 19 & 10 & 19 & 19 & 19 & 19 & 19 & 19 \\
		10 & 10 & 19 & 10 & 10 & 10 & 7 & 7 & 7 \\
		8 & 8 & 8 & 7 & 17 & 17 & 10 & 10 & 10 \\
		7 & 17 & 17 & 17 & 7 & 8 & 17 & 17 & 17 \\
		17 & 7 & 23 & 8 & 8 & 7 & 4 & 4 & 4 \\
		4 & 4 & 25 & 4 & 23 & 23 & 23 & 23 & 13 \\
		23 & 23 & 7 & 23 & 4 & 25 & 29 & 25 & 23 \\
		25 & 25 & 4 & 25 & 25 & 4 & 25 & 29 & 25 \\
		29 & 29 & 29 & 29 & 29 & 13 & 8 & 8 & 8 \\
		\bottomrule
	\end{tabular}
\end{table}

No significant change in WPR scores is observed for
different selections of $\theta$ and $\rho$. This is due to the
characteristic of PA rule that nodes of high in-degree (in-strength)
are likely to attract more incoming connections. When there is a
subset of nodes that have received the majority of incoming edges at
an early stage,
the newly generated edges will be connected towards these nodes with
high probabilities. While the in-strengths of these nodes keep
growing, their in-degrees increase as well, which results in high
scores of classical PR measure. Accordingly, the effect of edge 
weight on the final ranking results, especially the top~5, has 
become limited for PA networks. Rankings sensitive to $\theta$ are 
illustrated in the next example.

\subsection{Stochastic Block Model}
\label{sec:sbm}

\begin{figure}[tbp]
	\begin{center}
		\includegraphics[width = 0.68\textwidth]{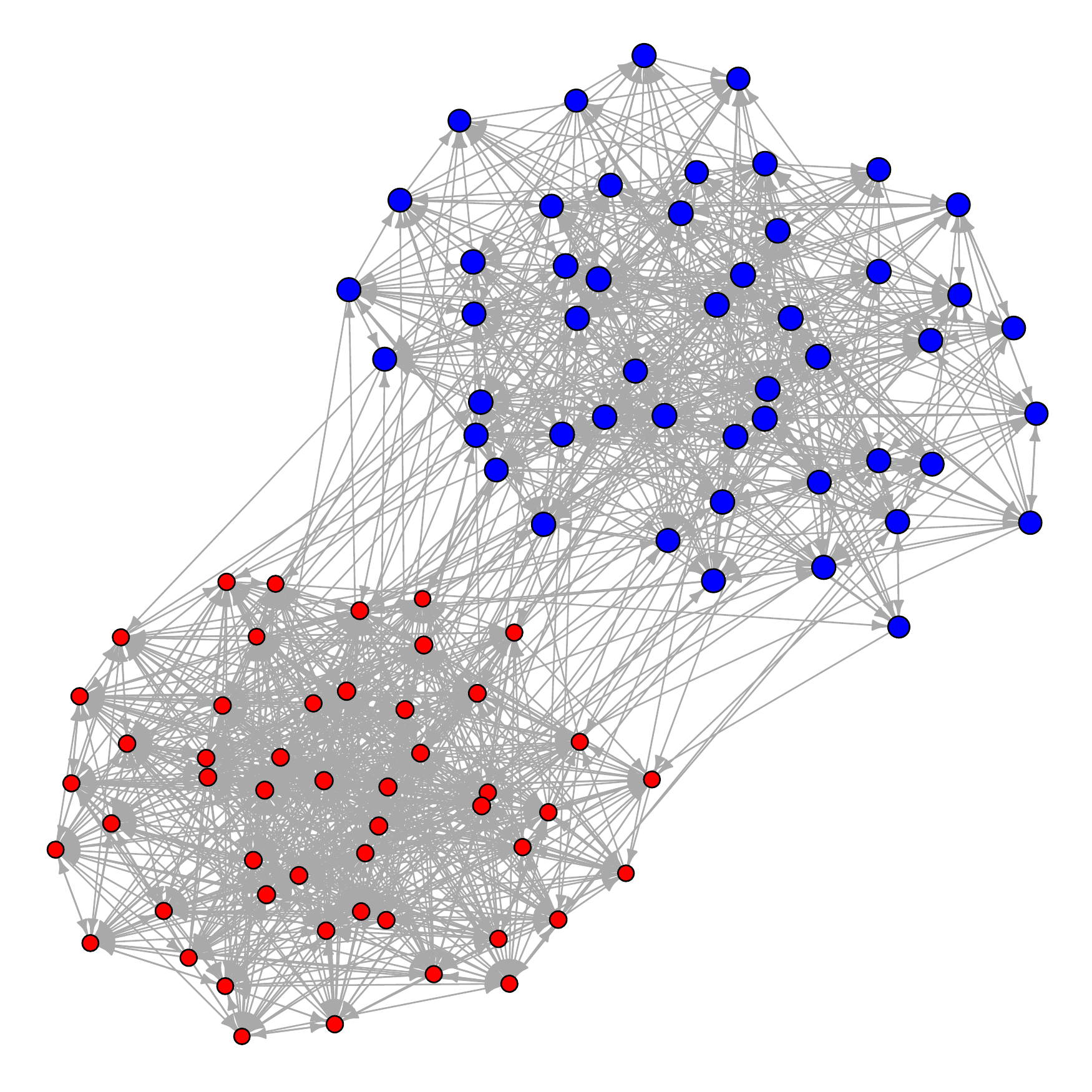}
	\end{center}
	\caption{A simulated (weighted and directed) SBM network
	consisting of two communities; the blue nodes are from $C_1$ and
	the red ones are from $C_2$; self-loops exist but are not
	presented.}
	\label{fig:SBM_ex}
\end{figure}

Stochastic block models (SBMs) are a class of network models for
characterizing community structure~\citep{holland1983stochastic,
snijders1997estimation, nowicki2001estimation}. In essence, an SBM
is comprised of a certain number of within-block
Erd\"{o}s-Renyi~\citep[ER,][]{erdos1959on}
models, where the cross-block structure is specified by Bernoulli
models. We generated a weighted and directed SBM network
consisting of two communities, $C_1$ and $C_2$, each containing $50$
members. The link densities within $C_1$ and $C_2$ were respectively
$0.2$ and $0.3$, whereas the link density between $C_1$ and $C_2$
was $0.02$. The weights for edges in $C_1$ and $C_2$ were
were independently drawn from ${\rm Bin}(500, 0.75)$ and
${\rm Bin}(20, 0.5)$, respectively.. The weights for edges $C_1$
and $C_2$ (either from $C_1$ to $C_2$ or from $C_2$ to $C_1$)
were independently drawn from ${\rm Bin}(5, 0.5)$. Community $C_1$ is
sparser than community $C_2$, but their within-community links
are both much denser than the between-community links.
In addition, edge weights in $C_1$ are much larger
than those in $C_2$, and edge weights between the communities
are the smallest. The generated SBM network is presented in
Figure~\ref{fig:SBM_ex}, where the node sizes are proportional to
the logarithm of their strengths. The nodes in $C_1$ and $C_2$ are
colored with blue and red, respectively.

Table~\ref{tab:SBM_sim} summarizes the top~10 nodes based on the WPR
scores with $\theta \in \{0, 0.5, 1\}$ and $\gamma = 0.85$.
When edge weight is not accounted for (i.e.,
$\theta = 0$), nine of the top~10 nodes come from $C_2$
since the nodes therein are more densely connected. A close
inspection shows that the top~3 nodes 59, 63 and 93 have the largest
in-degrees. For $\theta = 1$, all top~10 nodes belong to $C_1$. Node
39 has in-strength 2,230, which is less than that of node 50 (2,415),
but still ranks higher than node~50. This is desired as node 50
has a large amount of inputs from insignificant nodes,
like nodes 22 (rank 87), 23 (rank 100), 28
(rank 85) and 29 (rank 90), whereas node 39 has inputs mostly
from high rank nodes including itself. The top~3 nodes
in the list of $\theta = 0$ rank only, respectively,
56, 97 and 54, as they have low in-strengths, and the WPR scores
of the nodes linking towards them are relatively low.
For the hybrid case of $\theta = 0.5$, we observe a mixture of
the nodes from the
top~10 lists of $\theta = 0$ and $\theta = 1$ with five each. The
top~1 is node 59 (top 1 from $\theta = 0$), whose WPR score is
slightly higher than that of the second largest, node 39 (top 1 form
$\theta = 1$). The node with third highest WPR score is node 8,
which also comes from the $\theta = 1$ list. We see that the
difference between the WPR scores between nodes 39 and 8 is smaller
than that in the list of $\theta = 1$ since the edge weight is not
yet fully accounted.

\begin{table}[tbp]
	\centering
	\caption{The nodes with top 10 proposed WPR scores in the
	simulated SBM network for $\theta \in \{0, 0.5, 1\}$; the
	damping factor is fixed $\gamma = 0.85$.}
	\label{tab:SBM_sim}
	\setlength{\tabcolsep}{7pt}
	\begin{tabular}{cc cc cc}
		\toprule
		\multicolumn{2}{c}{$\phi$ ($\theta = 0$)} &
		\multicolumn{2}{c}{$\phi$ ($\theta = 0.5$)} &
		\multicolumn{2}{c}{$\phi$ ($\theta = 1$)} \\
		\cmidrule(lr){1-2} \cmidrule(lr){3-4}
		\cmidrule(lr){5-6}
		Node & WPR (\%) & Node & WPR (\%) & Node & WPR (\%) \\
		\midrule
		59 & 1.775 & 59 & 1.536 & 39 & 1.750 \\
		63 & 1.500 & 39 & 1.492 & 8 & 1.605 \\
		93 & 1.494 & 8 & 1.466 & 50 & 1.583 \\
		98 & 1.444 & 63 & 1.424 & 32 & 1.576 \\
		78 & 1.429 & 32 & 1.367 & 36 & 1.525 \\
		57 & 1.417 & 93 & 1.364 & 5 & 1.421 \\
		86 & 1.403 & 36 & 1.352 & 25 & 1.420 \\
		71 & 1.390 & 98 & 1.343 & 38 & 1.384 \\
		8 & 1.390 & 78 & 1.328 & 46 & 1.353 \\
		72 & 1.372 & 50 & 1.328 & 30 & 1.327 \\
		\bottomrule
	\end{tabular}
\end{table}

The SBM example provides strong evidence for the necessity of
accounting for edge weight in the WPR computation. Unlike the
example of PA network, the top~10 nodes here are
almost completely different between $\theta= 0$ and
$\theta = 1$. Noticeable changes have been observed in the WPR
scores as well. Such drastic changes in node ranks are primarily
due to the structure of the network. There is no preferential
attachment feature in the generation of SBMs, so the impact of edge
weight on the WPR scores remains compelling.

\section{World Input-Output Networks}
\label{sec:wion}

\begin{figure}[tbp]
	\begin{center}
		\includegraphics[width = 0.68\textwidth]{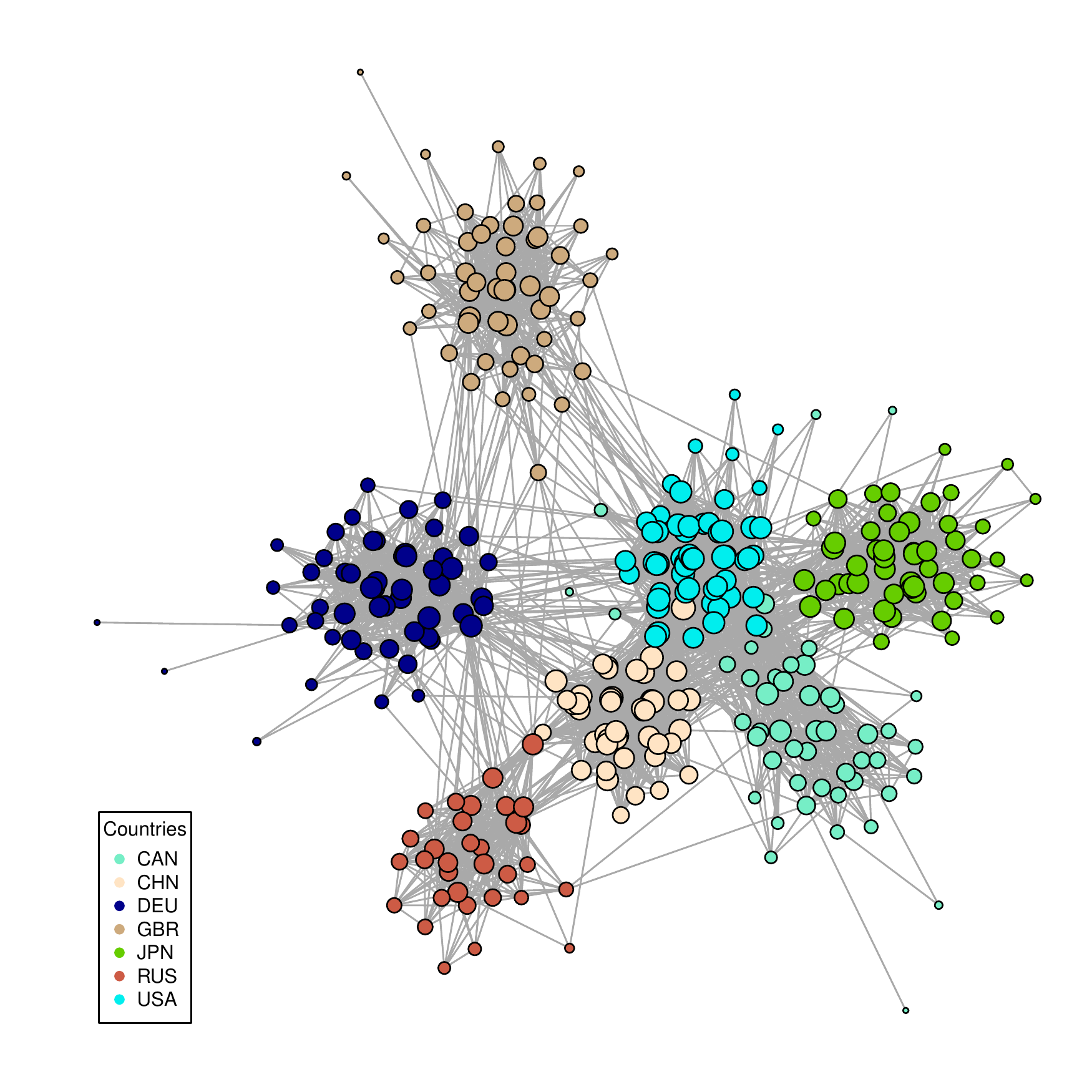}
	\end{center}
	\caption{A example of sub-network of the WION consisting of seven major
	economies in 2014; self-loops and isolated nodes have been
	removed; node sizes are proportional to the natural logarithm of
	their strengths; the edges of weight greater than or equal to
	$500$ (with unit 1 million USD) are presented.}
	\label{fig:net_ex_2014}
\end{figure}

In economics, a World Input-Output Table
(WIOT) is a multi-regional input-output table, which records the
intermediate transaction volumes among the sectors
from different countries/regions. It has great research value
in analyzing the inter-dependency across multi-regional sectors in
the global economy. 
We applied the proposed WPR to the WIONs constructed from the
annual WIOTs~\citep[WIOTs,][]{timmer2015anillustrated} from
2000 to 2014 using the 2016 release of the
\href{http://www.wiod.org/home}{World Input-Output Database}.
The 2016 release covers 56 sectors
from 44 countries/regions, including a region called ``the rest of
the world'' (ROW).  The dictionary for the sector codes are
given in Table~\ref{tab:code} in Appendix~\ref{sec:code}. The 2,464
region-sectors are the nodes of the WIONs. A transaction
from one region-sector to another forms a weighted, directed edge,
where the edge weight is represented by the transaction volume (in
the unit of 1 million USD). The
link densities of the WIONs are high with average $83\%$. A
sub-network consisting of seven major economies for 2014
is depicted in Figure~\ref{fig:net_ex_2014}. Only edges of
weight $\ge 500$ are presented, while the self-loops and
isolated nodes have been removed. The node size is proportional to
the natural logarithm of its total strength. A few studies have
investigated centrality measures (not limited to PR
and its variants) of the WIONs~\citep{cerina2015world,
del2017trends, xu2019input}.

\begin{table}[tbp]
	\centering
	\renewcommand{\arraystretch}{1.2}
	\caption{The region-sectors with top 10 WPR scores (no prior
		information) in the WIONs from 2000, 2007 and 2014, with
		$\gamma
		= 0.85$ and $\theta \in \{0,
		0.5, 1\}$.}
	\label{tab:wionwpr}
	\small \setlength{\tabcolsep}{5pt}
	\begin{tabular}{l ccc ccc ccc}
		\toprule
		\multirow{3}{*}{Rank} & \multicolumn{3}{c}{2000}
		& \multicolumn{3}{c}{2007} & \multicolumn{3}{c}{2014}
		\\ \cmidrule(lr){2-4} \cmidrule(lr){5-7} \cmidrule(lr){8-10}
		& $\theta = 0$ & $\theta = 0.5$ & $\theta = 1$ & $\theta =
		0$ & $\theta = 0.5$ & $\theta = 1$ & $\theta = 0$ & $\theta
		= 0.5$ & $\theta = 1$ \\
		\midrule
		1 & IND39 & ROW27 & USA51 & IND39 & ROW27 & ROW27 & IND39 &
		ROW27 & ROW27 \\
		2 & IND31 & DEU20 & ROW27 & IND31 & DEU20 & GBR53 & IND31 &
		ROW4 & USA51 \\
		3 & IND40 & USA51 & GBR53 & MEX5 & ROW4 & USA51 & IND40 &
		DEU20 & GBR53 \\
		4 & IND27 & ROW24 & USA53 & MEX50 & USA51 & USA53 & MEX50 &
		USA51 & CHN27 \\
		5 & IND32 & ROW6 & DEU20 & MEX27 & ROW24 & ROW24 & MEX5 &
		ROW24 & ROW4 \\
		6 & IND1 & ROW29 & USA20 & MEX28 & ESP27 & DEU20 & MEX27 &
		ROW5 & ROW24 \\
		7 & IND42 & USA27 & ROW6 & MEX29 & FRA27 & ESP27 & MEX30 &
		ROW51 & USA53 \\
		8 & MEX5 & DEU27 & USA27 & MEX30 & ROW6 & ROW4 & IND27 &
		GBR53 & ROW5 \\
		9 & MEX50 & USA20 & USA44 & IND40 & GBR53 & ROW6 & MEX29 &
		CHN27 & DEU20 \\
		10 & MEX27 & ROW31 & ROW24 & MEX45 & ROW17 & USA44 & MEX45 &
		ROW29 & ROW6 \\
		\bottomrule
	\end{tabular}
\end{table}

Table~\ref{tab:wionwpr} presents the top~10 region-sectors ranked
by the proposed WPR with $\theta \in \{0, 0.5, 1\}$ and $\gamma = 0.85$
for the WIONs from 2000, 2007 and 2014. When edge weights are not
taken into account (i.e. $\theta = 0$), all top~10 nodes are sectors from India
(IND) and Mexico (MEX) for all three years. These results are
counter-intuitive as neither of the two countries was regarded as the
most influential in the world economy during the study period. When
weight is partially accounted ($\theta = 0.5$), results are more
reasonable, but completely different from those with $\theta = 0$.
Construction (27) from ROW took the first place in all
three years. Manufacture of motor vehicles, trailers and
semi-trailers (20) from Germany (DEU) took the second place in 2000
and 2007, but the third place in 2014, and the second place in
2014 was taken by mining and quarrying (4) from ROW. Several other
traditional leading sectors from influential economies are also included in the
top~10 list, such as public administration and defense
and compulsory social security (51) from the United States of
America (USA) and human health and social work activities (53) from
the United Kingdom (GBR). Quite a few sectors from ROW besides
construction (27) are in the top~10 lists, which is understandable
since ROW aggregates over those outside of the 43
countries/regions.

When edge weight is fully accounted ($\theta = 1$), no significantly
different results have been observed from those with $\theta =0.5$.
Over all three years, the top~3 sectors were public administration and
defense and compulsory social security (51) from USA, human health and
social work activities (53) from GBR, and construction (27) from ROW,
except in different orders. Fewer sectors are from ROW in the top~10
lists of $\theta = 1$ compared to those of $\theta = 0.5$. As ROW
contains many countries/regions, its sectors have input and output
connections with all of the other region-sectors in the networks.
Those edge weights are not necessarily large even though they
are aggregated counts. Therefore, more leading sectors from the
strong economies appear in the top~10 lists.
In 2000, four sectors were from the world's largest economy USA,
but the number was reduced to three in 2007. Human health and social
work activities (53) and real estate activities (44) from USA were
indeed world-leading region-sectors. After the subprime
mortgage crisis in 2008, real estate activities
(44) did not appear in the top~10 list in 2014 like in 2000 or 2007.
Construction (27) from China (CHN) joined the top~10 list in 2014. As
China became the world's second largest economic power in 2010,
construction as a sector with the largest pulling effect 
has been expected to have a high rank~\citep{wang2021regional}. The
inclusion of a non European Union (EU) country or USA in the top~10
list suggests a changed landscape and increased diversity of the
global economy.

We next used the total value-added (TVA) in the WIOTs as prior
information in the WPR~\citep{wang2021regional}. The results
with edge weights fully accounted (i.e., $\theta = 1$) are
summarized in Table~\ref{tab:wionwpr_prior}.
Significant changes have been observed over time. In 2000, the top~9
region-sectors were from USA. The top~5 are public administration
and defense and compulsory social security (51), human health and
social work activities (53), real estate activities (44), construction
(27), and manufacture of motor vehicles, trailers and semi-trailers
(20). The only non-USA region-sector
was construction (27) from Japan, which has been a
large component of the Japanese economy in terms of output and
employment, and a robust force for the economic recovery and
expansion in Japan in the post-war years till 
today~\citep{bon1990historical, raftery1998globalization}.
In 2007, USA was not as dominant as in 2000 but still with six in the
top~10. The top~3 remain unchanged. Construction (27) from China
ranked the 9th. This result seems to be more reasonable, as
the Chinese government provided unlimited support to the
construction industry in 2007 in preparation for the 2008 Olympic
Game. As the most influential sector
in China, construction (27) has driven the development of a large
number of domestic sectors as well as international cooperation over
that period~\citep{broudehoux2007spectacular, zhang2007impact}. In 
2014, the top~10 nodes had 4 
from China, 3 from USA,
and 3 from ROW.  Construction (27) of China ranked the first. The other
three Chinese sectors were manufacture of motor
vehicles, trailers and semi-trailers (20), manufacture of computer,
electronic and optical products (17) and manufacture of basic metals
(15), which, respectively, ranked 6, 7 and 10. The top~3 USA sectors
in 2007 now ranked 2, 4, and 5. No EU sectors showed up in the top~10.
This result is consistent with the fact that USA and China are the two
largest economies in the world. Compared to results without TVA
prior, fewer region-sectors from ROW ranked in the top~10 as
their TVA amounts were small in general. 

\begin{table}[tbp]
	\centering
	\renewcommand{\arraystretch}{1.2}
	\caption{The region-sectors with top 10 WPR scores (with TVA
	prior information) in the WIONs from 2000, 2007 and 2014, with
		$\gamma	= 0.85$ and $\theta = 1$.}
	\label{tab:wionwpr_prior}
        \setlength{\tabcolsep}{5pt}
	\begin{tabular}{l cc cc cc}
		\toprule
		\multirow{3}{*}{Rank} & \multicolumn{2}{c}{2000}
		& \multicolumn{2}{c}{2007} & \multicolumn{2}{c}{2014}
		\\ \cmidrule(lr){2-3} \cmidrule(lr){4-5} \cmidrule(lr){6-7}
		& no prior & TVA prior & no prior & TVA prior & no prior &
		TVA prior \\
		\midrule
		1 & USA51 & USA51 & ROW27 & USA51 & ROW27 & CHN27 \\
		2 & ROW27 & USA53 & GBR53 & USA53 & USA51 & USA51 \\
		3 & GBR53 & USA44 & USA51 & USA44 & GBR53 & ROW27 \\
		4 & USA53 & USA27 & USA53 & ROW27 & CHN27 & USA53 \\
		5 & DEU20 & USA20 & ROW24 & USA27 & ROW4 & USA44 \\
		6 & USA20 & USA5 & DEU20 & GBR53 & ROW24 & CHN20 \\
		7 & ROW6 & USA30 & ESP27 & ROW24 & USA53 & CHN17 \\
		8 & USA27 & USA36 & ROW4 & USA5 & ROW5 & ROW4 \\
		9 & USA44 & USA29 & ROW6 & CHN27 & DEU20 & ROW24 \\
		10 & ROW24 & JPN27 & USA44 & USA29 & ROW6 & CHN15 \\
		\bottomrule
	\end{tabular}
\end{table}

\section{Discussions}
\label{sec:dis}

We propose a weighted PageRank measure that simultaneously accounts
for edge weights and prior information on the
relative importance of nodes in weighted directed networks.
The relative importance of node strengths and edge weights is controlled
by a tuning parameter for flexibility.
Efficient algorithms are implemented and made publicly available
in \proglang{R} package \proglang{wdnet} \citep{Rpkg:wdnet}.
Through two simulated network examples and one application to the WIONs, we have
observed significant differences between the results from the
proposed WPR and other classical measures, where the proposed WPR is
preferred. Especially for the WIONs, the proposed WPR has led to
conclusions that are more consistent with intuition, providing new insights into the
global input-output system. Both synthetic and real data studies
suggest the need for considering edge weight and prior
information in the node centrality measure for weighted
directed networks.

There are several limitations in the present research that merit further
studies. So far the proposed measure has been adapted to static
networks only. It is of
substantial interest to investigate the proposed measure in random
network models. Such extension would provide theoretical foundations for
statistical inference such as confidence interval and hypothesis
testing. Recently, \citet{avrachenkov2015pagerank} and
\citet{banerjee2021pagerank} have looked into
the asymptotic properties of the classical PR in undirected,
unweighted SBMs and directed, unweighted PA networks, respectively.
In addition to only being a centrality measure, the classical PR has been
used to identify community structure in unweighted
networks~\citep{kloumann2017block}. Applying the proposed
WPR in community detection to weighted networks may lead to fruitful
results.

\bibliographystyle{model1-num-names}

\begin{thebibliography}{39}
\expandafter\ifx\csname natexlab\endcsname\relax\def\natexlab#1{#1}\fi
\providecommand{\url}[1]{\texttt{#1}}
\providecommand{\href}[2]{#2}
\providecommand{\path}[1]{#1}
\providecommand{\DOIprefix}{doi:}
\providecommand{\ArXivprefix}{arXiv:}
\providecommand{\URLprefix}{URL: }
\providecommand{\Pubmedprefix}{pmid:}
\providecommand{\doi}[1]{\href{http://dx.doi.org/#1}{\path{#1}}}
\providecommand{\Pubmed}[1]{\href{pmid:#1}{\path{#1}}}
\providecommand{\bibinfo}[2]{#2}
\ifx\xfnm\relax \def\xfnm[#1]{\unskip,\space#1}\fi
\bibitem[{Freeman(1978)}]{freeman1978centrality}
\bibinfo{author}{L.~C. Freeman},
\newblock \bibinfo{title}{Centrality in social networks conceptual
  clarification},
\newblock \bibinfo{journal}{Social Networks} \bibinfo{volume}{1}
  (\bibinfo{year}{1978}) \bibinfo{pages}{215--239}.
\bibitem[{Newman(2001)}]{newman2001scientific}
\bibinfo{author}{M.~E.~J. Newman},
\newblock \bibinfo{title}{Scientific collaboration networks. {II}. {S}hortest
  paths, weighted networks, and centrality},
\newblock \bibinfo{journal}{Physical Review E} \bibinfo{volume}{64}
  (\bibinfo{year}{2001}) \bibinfo{pages}{016132}.
\bibitem[{Bonacich(1972)}]{bonacich1972factoring}
\bibinfo{author}{P.~Bonacich},
\newblock \bibinfo{title}{Factoring and weighting approaches to status scores
  and clique identification},
\newblock \bibinfo{journal}{The Journal of Mathematical Sociology}
  \bibinfo{volume}{2} (\bibinfo{year}{1972}) \bibinfo{pages}{113--120}.
\bibitem[{Katz(1953)}]{katz1953anew}
\bibinfo{author}{L.~Katz},
\newblock \bibinfo{title}{A new status index derived from sociometric
  analysis},
\newblock \bibinfo{journal}{Psychometrika} \bibinfo{volume}{18}
  (\bibinfo{year}{1953}) \bibinfo{pages}{39--43}.
\bibitem[{Page et~al.(1999)Page, Brin, Motwani, and
  Winograd}]{brin1999thepagerank}
\bibinfo{author}{L.~Page}, \bibinfo{author}{S.~Brin},
  \bibinfo{author}{R.~Motwani}, \bibinfo{author}{T.~Winograd},
  \bibinfo{title}{The {P}age{R}ank Citation Ranking: {B}ringing {O}rder to the
  Web}, \bibinfo{type}{Technical Report} \bibinfo{number}{1999-66}, Stanford
  University, \bibinfo{year}{1999}.
\bibitem[{Wang et~al.(2011)Wang, Mo, Wang, and Jin}]{wang2011exploring}
\bibinfo{author}{J.~Wang}, \bibinfo{author}{H.~Mo}, \bibinfo{author}{F.~Wang},
  \bibinfo{author}{F.~Jin},
\newblock \bibinfo{title}{Exploring the network structure and nodal centrality
  of {C}hina’s air transport network: {A} complex network approach},
\newblock \bibinfo{journal}{Journal of Transport Geography}
  \bibinfo{volume}{19} (\bibinfo{year}{2011}) \bibinfo{pages}{712--721}.
\bibitem[{Hirsch(2005)}]{hirsch2005index}
\bibinfo{author}{J.~E. Hirsch},
\newblock \bibinfo{title}{An index to quantify an individual's scientific
  research output},
\newblock \bibinfo{journal}{Proceedings of the National academy of Sciences of
  the United States of America} \bibinfo{volume}{102} (\bibinfo{year}{2005})
  \bibinfo{pages}{16569--16572}.
\bibitem[{Das et~al.(2018)Das, Smanta, and Pal}]{das2018study}
\bibinfo{author}{K.~Das}, \bibinfo{author}{S.~Smanta},
  \bibinfo{author}{M.~Pal},
\newblock \bibinfo{title}{Study on centrality measures in social networks: {A}
  survey},
\newblock \bibinfo{journal}{Social Network Analysis and Mining}
  \bibinfo{volume}{8} (\bibinfo{year}{2018}) \bibinfo{pages}{8}.
\bibitem[{Newman(2018)}]{newman2018networks}
\bibinfo{author}{M.~E.~J. Newman}, \bibinfo{title}{Networks},
  \bibinfo{publisher}{Oxford University Press}, \bibinfo{address}{Oxford, UK},
  \bibinfo{year}{2018}.
\bibitem[{Brin and Page(1998)}]{brin1998theanatomy}
\bibinfo{author}{S.~Brin}, \bibinfo{author}{L.~Page},
\newblock \bibinfo{title}{The anatomy of a large-scale hypertextual web search
  engine},
\newblock \bibinfo{journal}{Computer Networks and ISDN Systems}
  \bibinfo{volume}{30} (\bibinfo{year}{1998}) \bibinfo{pages}{107--117}.
\bibitem[{Yan and Ding(2011)}]{yan2011discovering}
\bibinfo{author}{E.~Yan}, \bibinfo{author}{Y.~Ding},
\newblock \bibinfo{title}{Discovering author impact: {A} {P}age{R}ank
  perspective},
\newblock \bibinfo{journal}{Information Processing \& Management}
  \bibinfo{volume}{47} (\bibinfo{year}{2011}) \bibinfo{pages}{125--134}.
\bibitem[{Ding(2011)}]{ding2011applying}
\bibinfo{author}{Y.~Ding},
\newblock \bibinfo{title}{Applying weighted {P}age{R}ank to author citation
  network},
\newblock \bibinfo{journal}{Journal of the American Society for Information
  Science and Technology} \bibinfo{volume}{62} (\bibinfo{year}{2011})
  \bibinfo{pages}{236--245}.
\bibitem[{Kalecky and Cho(2018)}]{kalecky2018primalign}
\bibinfo{author}{K.~Kalecky}, \bibinfo{author}{Y.-R. Cho},
\newblock \bibinfo{title}{Prim{A}lign: {P}age{R}ank-inspired {M}arkovian
  alignment for large biological networks},
\newblock \bibinfo{journal}{Bioinformatics} \bibinfo{volume}{34}
  (\bibinfo{year}{2018}) \bibinfo{pages}{537--546}.
\bibitem[{Newman(2004)}]{newman2004analysis}
\bibinfo{author}{M.~E.~J. Newman},
\newblock \bibinfo{title}{Analysis of weighted networks},
\newblock \bibinfo{journal}{Physical Review E} \bibinfo{volume}{70}
  (\bibinfo{year}{2004}) \bibinfo{pages}{056131}.
\bibitem[{Xing and Ghorbani(2004)}]{xing2004weighted}
\bibinfo{author}{W.~Xing}, \bibinfo{author}{A.~A. Ghorbani},
\newblock \bibinfo{title}{Weighted {P}age{R}ank algorithm},
\newblock in: \bibinfo{editor}{A.~A. Ghorbani} (Ed.),
  \bibinfo{booktitle}{Proceedings of the 2nd Annual Conference on Communication
  Networks and Services Research}, \bibinfo{publisher}{IEEE},
  \bibinfo{address}{Piscataway, NJ, USA}, \bibinfo{year}{2004}, pp.
  \bibinfo{pages}{305--314}.
\bibitem[{Timmer et~al.(2015)Timmer, Dietzenbacher, Los, Stehrer, and
  de~Vries}]{timmer2015anillustrated}
\bibinfo{author}{M.~P. Timmer}, \bibinfo{author}{E.~Dietzenbacher},
  \bibinfo{author}{B.~Los}, \bibinfo{author}{R.~Stehrer},
  \bibinfo{author}{G.~J. de~Vries},
\newblock \bibinfo{title}{An illustrated user guide to the {W}orld
  {I}nput-{O}utput {D}atabase: {T}he case of global automotive production},
\newblock \bibinfo{journal}{Review of International Economics}
  \bibinfo{volume}{23} (\bibinfo{year}{2015}) \bibinfo{pages}{575--605}.
\bibitem[{Yan et~al.(2020)Yan, Yuan, and Zhang}]{Rpkg:wdnet}
\bibinfo{author}{J.~Yan}, \bibinfo{author}{Y.~Yuan},
  \bibinfo{author}{P.~Zhang}, \bibinfo{title}{{wdnet}: {W}eighted Directed
  Network}, \bibinfo{organization}{University of Connecticut},
  \bibinfo{year}{2020}. \bibinfo{note}{{R} package version 0.0-3,
  \url{https://gitlab.com/wdnetwork/wdnet}}.
\bibitem[{Berkhin(2005)}]{berkhin2005asurvey}
\bibinfo{author}{P.~Berkhin},
\newblock \bibinfo{title}{A survey on {P}age{R}ank computing},
\newblock \bibinfo{journal}{Internet Mathematics} \bibinfo{volume}{2}
  (\bibinfo{year}{2005}) \bibinfo{pages}{73--120}.
\bibitem[{Perron(1907)}]{perron1907zur}
\bibinfo{author}{O.~Perron},
\newblock \bibinfo{title}{Zur theorie der matrices},
\newblock \bibinfo{journal}{Mathematische Annalen} \bibinfo{volume}{64}
  (\bibinfo{year}{1907}) \bibinfo{pages}{248--263}.
\bibitem[{Lehoucq et~al.(1998)Lehoucq, Sorensen, and Yang}]{lehoucq1998arpack}
\bibinfo{author}{R.~B. Lehoucq}, \bibinfo{author}{D.~C. Sorensen},
  \bibinfo{author}{C.~Yang}, \bibinfo{title}{ARPACK Users' Guide: {S}olution of
  Large-Scale Eigenvalue Problems with Implicitly Restarted {A}rnoldi Methods},
  \bibinfo{publisher}{SIAM}, \bibinfo{address}{Philadelphia, PA, USAc},
  \bibinfo{year}{1998}.
\bibitem[{Qiu and Mei(2016)}]{qiu2016rARPACK}
\bibinfo{author}{Y.~Qiu}, \bibinfo{author}{J.~Mei}, \bibinfo{title}{rARPACK:
  Solvers for Large Scale Eigenvalue and SVD Problems}, \bibinfo{year}{2016}.
  \bibinfo{note}{{R} package version 0.11-0,
  \url{https://CRAN.R-project.org/package=rARPACK}}.
\bibitem[{Barab\'{a}si and Albert(1999)}]{barabasi1999emergence}
\bibinfo{author}{A.-L. Barab\'{a}si}, \bibinfo{author}{R.~Albert},
\newblock \bibinfo{title}{Emergence of scaling in random networks},
\newblock \bibinfo{journal}{Science} \bibinfo{volume}{286}
  (\bibinfo{year}{1999}) \bibinfo{pages}{509--512}.
\bibitem[{Yuan et~al.(2021)Yuan, Yan, and Zhang}]{yuan2021assortativity}
\bibinfo{author}{Y.~Yuan}, \bibinfo{author}{J.~Yan},
  \bibinfo{author}{P.~Zhang}, \bibinfo{title}{Assortativity measures for
  weighted and directed networks}, \bibinfo{year}{2021}.
  \href{http://arxiv.org/abs/arXiv:2101.05389}{\tt arXiv:arXiv:2101.05389},
  \bibinfo{note}{\href{https://arxiv.org/pdf/2101.05389.pdf}{arXiv:2101.05389}}.
\bibitem[{Wang and Zhang(2021)}]{wang2021directed}
\bibinfo{author}{T.~Wang}, \bibinfo{author}{P.~Zhang}, \bibinfo{title}{Directed
  hybrid random networks mixing preferential attachment with uniform attachment
  mechanisms}, \bibinfo{year}{2021}.
  \href{http://arxiv.org/abs/arXiv:2101.04611}{\tt arXiv:arXiv:2101.04611},
  \bibinfo{note}{\href{https://arxiv.org/pdf/2101.04611.pdf}{arXiv:2101.04611}}.
\bibitem[{Holland et~al.(1983)Holland, Laskey, and
  Leinhadrt}]{holland1983stochastic}
\bibinfo{author}{P.~W. Holland}, \bibinfo{author}{K.~B. Laskey},
  \bibinfo{author}{S.~Leinhadrt},
\newblock \bibinfo{title}{Stochastic blockmodels: First steps},
\newblock \bibinfo{journal}{Social Networks} \bibinfo{volume}{5}
  (\bibinfo{year}{1983}) \bibinfo{pages}{109--137}.
\bibitem[{Snijders and Nowicki(1997)}]{snijders1997estimation}
\bibinfo{author}{T.~A.~B. Snijders}, \bibinfo{author}{K.~Nowicki},
\newblock \bibinfo{title}{Estimation and prediction for stochastic blockmodels
  for graphs with latent block structure},
\newblock \bibinfo{journal}{Journal of Classification} \bibinfo{volume}{14}
  (\bibinfo{year}{1997}) \bibinfo{pages}{75--100}.
\bibitem[{Nowicki and Snijders(2001)}]{nowicki2001estimation}
\bibinfo{author}{K.~Nowicki}, \bibinfo{author}{T.~A.~B. Snijders},
\newblock \bibinfo{title}{Estimation and prediction for stochastic
  blockstructures},
\newblock \bibinfo{journal}{Journal of the American Statistical Association}
  \bibinfo{volume}{96} (\bibinfo{year}{2001}) \bibinfo{pages}{1077--1087}.
\bibitem[{Erd\"{o}s and R\'{e}nyi(1959)}]{erdos1959on}
\bibinfo{author}{P.~Erd\"{o}s}, \bibinfo{author}{A.~R\'{e}nyi},
\newblock \bibinfo{title}{On random graphs i},
\newblock \bibinfo{journal}{Publicationes Mathematicae Debrecen}
  \bibinfo{volume}{6} (\bibinfo{year}{1959}) \bibinfo{pages}{290--297}.
\bibitem[{Cerina et~al.(2015)Cerina, Zhu, Chessa, and
  Riccaboni}]{cerina2015world}
\bibinfo{author}{F.~Cerina}, \bibinfo{author}{Z.~Zhu},
  \bibinfo{author}{A.~Chessa}, \bibinfo{author}{M.~Riccaboni},
\newblock \bibinfo{title}{World input-output network},
\newblock \bibinfo{journal}{PLOS One} \bibinfo{volume}{10}
  (\bibinfo{year}{2015}) \bibinfo{pages}{e0134025}.
\bibitem[{del R\'{i}o-Chanona et~al.(2017)del R\'{i}o-Chanona, Gruji\'{c}, and
  Jensen}]{del2017trends}
\bibinfo{author}{R.~M. del R\'{i}o-Chanona}, \bibinfo{author}{J.~Gruji\'{c}},
  \bibinfo{author}{H.~J. Jensen},
\newblock \bibinfo{title}{Trends of the world input and output network of
  global trade},
\newblock \bibinfo{journal}{PLOS One} \bibinfo{volume}{12}
  (\bibinfo{year}{2017}) \bibinfo{pages}{e0170817}.
\bibitem[{Xu and Sai(2019)}]{xu2019input}
\bibinfo{author}{M.~Xu}, \bibinfo{author}{L.~Sai},
\newblock \bibinfo{title}{Input-output networks offer new insights of economic
  structure},
\newblock \bibinfo{journal}{Physica A: Statistical Mechanics and its
  Applications} \bibinfo{volume}{527} (\bibinfo{year}{2019})
  \bibinfo{pages}{121178}.
\bibitem[{Wang et~al.(2021)Wang, Xiao, Yan, and Zhang}]{wang2021regional}
\bibinfo{author}{T.~Wang}, \bibinfo{author}{S.~Xiao}, \bibinfo{author}{J.~Yan},
  \bibinfo{author}{P.~Zhang}, \bibinfo{title}{Regional and sectoral structures
  and their dynamics of chinese economy: A network perspective from
  multi-regional input-output tables}, \bibinfo{year}{2021}.
  \href{http://arxiv.org/abs/arXiv:2102.12454}{\tt arXiv:arXiv:2102.12454},
  \bibinfo{note}{\href{https://arxiv.org/pdf/2102.12454.pdf}{arXiv:2102.12454}}.
\bibitem[{Bon and Pietroforte(1990)}]{bon1990historical}
\bibinfo{author}{R.~Bon}, \bibinfo{author}{R.~Pietroforte},
\newblock \bibinfo{title}{Historical comparison of construction sectors in the
  {U}nited {S}tates, {J}apan, {I}taly and {F}inland using input-output tables},
\newblock \bibinfo{journal}{Construction Management and Economics}
  \bibinfo{volume}{8} (\bibinfo{year}{1990}) \bibinfo{pages}{233--247}.
\bibitem[{Raftery et~al.(1998)Raftery, Pasadilla, Chiang, Hui, and
  Tang}]{raftery1998globalization}
\bibinfo{author}{J.~Raftery}, \bibinfo{author}{B.~Pasadilla},
  \bibinfo{author}{Y.~H. Chiang}, \bibinfo{author}{E.~C. Hui},
  \bibinfo{author}{B.-S. Tang},
\newblock \bibinfo{title}{Globalization and construction industry development:
  {I}mplications of recent developments in the construction sector in {A}sia},
\newblock \bibinfo{journal}{Construction Management and Economics}
  \bibinfo{volume}{16} (\bibinfo{year}{1998}) \bibinfo{pages}{729--737}.
\bibitem[{Broudehoux(2007)}]{broudehoux2007spectacular}
\bibinfo{author}{A.-M. Broudehoux},
\newblock \bibinfo{title}{Spectacular {B}eijing: {T}he conspicuous construction
  of an {O}lympic metropolis},
\newblock \bibinfo{journal}{Journal of Urban Affairs} \bibinfo{volume}{29}
  (\bibinfo{year}{2007}) \bibinfo{pages}{383--399}.
\bibitem[{Zhang and Zhao(2007)}]{zhang2007impact}
\bibinfo{author}{Y.~Zhang}, \bibinfo{author}{K.~Zhao},
\newblock \bibinfo{title}{Impact of {B}eijing {O}lympic‐related investments
  on regional economic growth of {C}hina: {I}nterregional input-output
  approach},
\newblock \bibinfo{journal}{Asian Economic Journal} \bibinfo{volume}{21}
  (\bibinfo{year}{2007}) \bibinfo{pages}{261--282}.
\bibitem[{Avrachenkov et~al.(2015)Avrachenkov, Kadavankandy, Prokhorenkov, and
  Raigorodskii}]{avrachenkov2015pagerank}
\bibinfo{author}{K.~Avrachenkov}, \bibinfo{author}{A.~Kadavankandy},
  \bibinfo{author}{L.~O. Prokhorenkov}, \bibinfo{author}{A.~Raigorodskii},
\newblock \bibinfo{title}{Page{R}ank in undirected random graphs},
\newblock in: \bibinfo{editor}{D.~Gleich}, \bibinfo{editor}{J.~Komj\'{a}thy},
  \bibinfo{editor}{N.~Litvak} (Eds.), \bibinfo{booktitle}{Proceedings of the
  International Workshop on Algorithms and Models for the Web-Graph (WAW
  2015)}, \bibinfo{publisher}{Springer}, \bibinfo{address}{Cham, Switzerland},
  \bibinfo{year}{2015}, pp. \bibinfo{pages}{151--163}.
\bibitem[{Banerjee and Olvera-Cravioto(2021)}]{banerjee2021pagerank}
\bibinfo{author}{S.~Banerjee}, \bibinfo{author}{M.~Olvera-Cravioto},
  \bibinfo{title}{Page{R}ank asymptotics on directed preferential attachment
  networks}, \bibinfo{year}{2021}.
  \href{http://arxiv.org/abs/arXiv:2102.08894}{\tt arXiv:arXiv:2102.08894},
  \bibinfo{note}{\href{https://arxiv.org/pdf/2102.08894.pdf}{arXiv:2102.08894}}.
\bibitem[{Kloumann et~al.(2017)Kloumann, Ugander, and
  Kleinberg}]{kloumann2017block}
\bibinfo{author}{I.~M. Kloumann}, \bibinfo{author}{J.~Ugander},
  \bibinfo{author}{J.~Kleinberg},
\newblock \bibinfo{title}{Block models and personalized {P}age{R}ank},
\newblock \bibinfo{journal}{Proceedings of the National academy of Sciences of
  the United States of America} \bibinfo{volume}{114} (\bibinfo{year}{2017})
  \bibinfo{pages}{33--38}.

\end{thebibliography}

\appendix

\section{Weighted PR by~\citet{xing2004weighted}}
\label{sec:xing}
\begin{figure}[tbp]
	\begin{center}
		\begin{tikzpicture}
			\draw (0,1) node[draw = black, circle, minimum size =
			0.86cm] (u1)	{$u_1$} ;
			\draw (0,3) node[draw = black, circle, minimum size =
			0.86cm] (u2) {$u_2$} ;
			\draw (3,0) node[draw = black, circle, minimum size =
			0.86cm] (v1) {$v_1$} ;
			\draw (3,2) node[draw = black, circle, minimum size =
			0.86cm] (v2) {$v_2$} ;
			\draw (3,4) node[draw = black, circle, minimum size =
			0.86cm] (v3) {$v_3$} ;
			\draw (6,0) node[draw = black, circle, minimum size =
			0.86cm] (w1) {$w_1$} ;
			\draw (6,2) node[draw = black, circle, minimum size =
			0.86cm] (w2) {$w_2$} ;
			\draw (6,4) node[draw = black, circle, minimum size =
			0.86cm] (w3) {$w_3$} ;
			\draw[-latex, thick] (u1) -- (v1) ;
			\draw[-latex, thick] (u1) -- (v2) ;
			\draw[-latex, thick] (u2) -- (v2) ;
			\draw[-latex, thick] (u2) -- (v3) ;
			\draw[-latex, thick] (v1) -- (w1) ;
			\draw[-latex, thick] (v1) -- (w3) ;
			\draw[-latex, thick] (v2) -- (w1) ;
			\draw[-latex, thick] (v2) -- (w2) ;
			\draw[-latex, thick] (v2) -- (w3) ;
			\draw[-latex, thick] (v3) -- (w2) ;
			\draw[-latex, thick] (v3) -- (w3) ;
		\end{tikzpicture}
	\end{center}
	\caption{A toy example for illustrating the edge weighting
	process in~\citet{xing2004weighted}; all the edges have a unit
	weight.}
	\label{fig:xingtoy}
\end{figure}
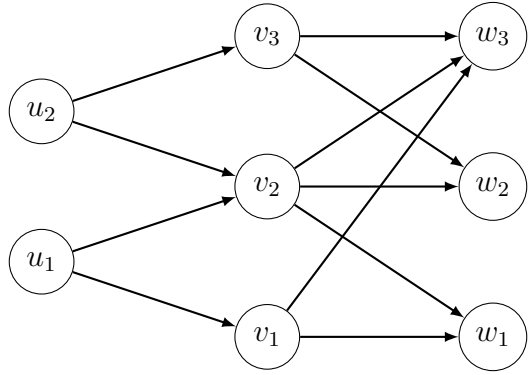

\citet{xing2004weighted} proposed a weighted PR to rank the web
pages based on their popularity, where the popularity of a web page
is reflected in two aspects: a large number of web pages have links
to it and a large number of web pages it is linked to. Xing and
Ghorbanis' weighted PR measure, after normalization, is
\begin{equation}
 	\label{eq:xing}
 	\phi^{\rm XG}(i) = \gamma \sum_{j \in V} a_{ji}
 	\left(\frac{\indeg_i}{\sum_{k \in V}
 	a_{jk} \indeg_k}\right) \left(\frac{\outdeg_i}{\sum_{k \in V}
 	a_{jk} \outdeg_k}\right) \phi^{\rm XG}(j)
 	+ \frac{1 - \gamma}{n}.
 \end{equation}
One of the distinctive features of this weighted PR measure is that
the edges of an unweighted network are actually ``weighted''. The
``weight'' of an edge is calculated by accounting for not only the
in-degree and out-degree of the target node of the edge, but also
the in-degrees and out-degrees of all the nodes that are linked by
the source node of the edge. To clarify, let us take the edge from
$u_1$ to $v_2$ in Figure~\ref{fig:xingtoy} as an example. Its
in-degree and out-degree generated weight are respectively given by
\[
\frac{\indeg_{v_2}}{\indeg_{v_1} + \indeg_{v_2}} = \frac{2}{1 + 2} =
\frac{2}{3} \qquad
\text{and} \qquad
\frac{\outdeg_{v_2}}{\outdeg_{v_1} + \outdeg_{v_2}} = \frac{3}{2 +
3}
= \frac{3}{5}.
\]
Xing and Ghorbanis' weighted PR does not actually use the
information quantified by edge weight; instead, the ``weight'' is
converted from node in- and out-degrees. Thus, the so-called
``weight'' is always integer-valued, causing the loss of generality.
In addition, this PR
measure does not seem to be applicable to some social networks. A
celebrity may follow only a few users in a social media
platform, so is likely to get a low score for Xing and Ghorbanis' PR
measure.

\section{Code Dictionary of the WIOTs}
\label{sec:code}

Table~\ref{tab:code} summarizes the code and definition of the 56
sectors in the 2016 release of the WIOD.

\begin{small}
  \singlespacing
	\begin{longtable}[c]{c p{0.9\textwidth}@{}}
          \caption{Description of the codes in the WIOTs}
          \label{tab:code}
          \endfirsthead
          \toprule
		Code & Sector \\
		\midrule
		1 & Crop and animal production, hunting and related service
		activities \\
		2 & Forestry and logging \\
		3 & Fishing and aquaculture \\
		4 & Mining and quarrying \\
		5 & Manufacture of food products, beverages and tobacco
		products \\
		6 &	Manufacture of textiles, wearing apparel and leather
		products \\
		7 &	Manufacture of wood and of products of wood and cork,
		except furniture; manufacture of articles of straw and
		plaiting materials \\
		8 & Manufacture of paper and paper products \\
		9 & Printing and reproduction of recorded media \\
		10 & Manufacture of coke and refined petroleum products \\
		11 & Manufacture of chemicals and chemical products \\
		12 & Manufacture of basic pharmaceutical products and
		pharmaceutical preparations \\
		13 & Manufacture of rubber and plastic products \\
		14 & Manufacture of other non-metallic mineral products \\
		15 & Manufacture of basic metals \\
		16 & Manufacture of fabricated metal products, except
		machinery and equipment \\
		17 & Manufacture of computer, electronic and optical
		products \\
		18 & Manufacture of electrical equipment \\
		19 & Manufacture of machinery and equipment n.e.c. \\
		20 & Manufacture of motor vehicles, trailers and
		semi-trailers \\
		21 & Manufacture of other transport equipment \\
		22 & Manufacture of furniture; other manufacturing \\
		23 & Repair and installation of machinery and equipment \\
		24 & Electricity, gas, steam and air conditioning supply \\
		25 & Water collection, treatment and supply \\
		26 & Sewerage; waste collection, treatment and disposal
		activities; materials recovery; remediation activities and
		other waste management services \\
		27 & Construction \\
		28 & Wholesale and retail trade and repair of motor vehicles
		and motorcycles \\
		29 & Wholesale trade, except of motor vehicles and
		motorcycles \\
		30 & Retail trade, except of motor vehicles and motorcycles
		\\
		31 & Land transport and transport via pipelines \\
		32 & Water transport \\
		33 & Air transport \\
		34 & Warehousing and support activities for transportation \\
		35 & Postal and courier activities \\
		36 & Accommodation and food service activities \\
		37 & Publishing activities \\
		38 & Motion picture, video and television program
		production, sound recording and music publishing activities;
		programming and	broadcasting activities \\
		39 & Telecommunications \\
		40 & Computer programming, consultancy and related
		activities; information service activities \\
		41 & Financial service activities, except insurance and
		pension funding \\
		42 & Insurance, reinsurance and pension funding, except
		compulsory social security \\
		43 & Activities auxiliary to financial services and
		insurance activities \\
		44 & Real estate activities \\
		45 & Legal and accounting activities; activities of head
		offices; management consultancy activities \\
		46 & Architectural and engineering activities; technical
		testing and	analysis \\
		47 & Scientific research and development \\
		48 & Advertising and market research \\
		49 & Other professional, scientific and technical
		activities; veterinary activities \\
		50 & Administrative and support service activities \\
		51 & Public administration and defense; compulsory social
		security \\
		52 & Education \\
		53 & Human health and social work activities \\
		54 & Other service activities \\
		55 & Activities of households as employers; undifferentiated
		goods- and services-producing activities of households for
		own use \\
		56 & Activities of extraterritorial organizations and bodies
		\\
		\bottomrule
	\end{longtable}
\end{small}

\end{document}